\documentclass[conference,10pt]{IEEEtran}
\usepackage{epsfig,epsf,rotating,setspace,latexsym,amsmath,amssymb,amsfonts,bm,theorem,subfigure,epstopdf}
\usepackage{cite,authblk}
\usepackage{bbm}
\usepackage{algorithm}
\usepackage[noend]{algpseudocode}
\usepackage{color}
\usepackage{mathtools}
\usepackage{soul}

\algrenewcommand\algorithmicforall{\textbf{foreach}}
\algrenewcommand\algorithmicindent{.8em}

\newtheorem{theorem}{Theorem}
\newtheorem{lemma}{Lemma}

\newenvironment{Proof}[1]{\medskip\par\noindent{\bf Proof:\,}\,#1}{{\mbox{\,$\blacksquare$}\par}}

\IEEEoverridecommandlockouts
\allowdisplaybreaks

\begin{document}
 
\title{ASUMAN: Age Sense Updating Multiple \\ Access in Networks}
 
\author{Purbesh Mitra \qquad Sennur Ulukus\\
        \normalsize Department of Electrical and Computer Engineering\\
        \normalsize University of Maryland, College Park, MD 20742\\
        \normalsize  \emph{pmitra@umd.edu} \qquad \emph{ulukus@umd.edu}}
\maketitle

\begin{abstract}
We consider a fully-connected wireless gossip network which consists of a source and $n$ receiver nodes. The source updates itself with a Poisson process and also sends updates to the nodes as Poisson arrivals. Upon receiving the updates, the nodes update their knowledge about the source. The nodes gossip the data among themselves in the form of Poisson arrivals to disperse their knowledge about the source. The total gossiping rate is bounded by a constraint. The goal of the network is to be as timely as possible with the source. In this work, we propose ASUMAN, a distributed opportunistic gossiping scheme, where after each time the source updates itself, each node waits for a time proportional to its current age and broadcasts a signal to the other nodes of the network. This allows the nodes in the network which have higher age to remain silent and only the low-age nodes to gossip, thus utilizing a significant portion of the constrained total gossip rate. We calculate the average age for a typical node in such a network with symmetric settings and show that the theoretical upper bound on the age scales as $O(1)$. ASUMAN, with an average age of $O(1)$, offers significant gains compared to a system where the nodes just gossip blindly with a fixed update rate in which case the age scales as $O(\log n)$. 
\end{abstract}

\section{Introduction}

Gossiping is a mechanism to disperse information quickly in a network. Each node of the network transmits its own data randomly to its neighboring nodes. This kind of technique is particularly useful in dense distributed sensor networks where a large number of nodes communicate with each other without the presence of a centralized server that schedules transmissions. Although gossiping has been studied extensively \cite{yaron03thesis, shah08monograph, Sanghavi2007GossipFileSplit}, the timeliness of gossiping networks is first analyzed in \cite{yates21gossip}. For measuring the timeliness of a system, the age of information metric has been introduced \cite{kaul11AoI, kosta17AoIbook, Sun2019AgeOI, yatesJSACsurvey}. A disadvantage of the traditional age metric is that it does not take the source's update rate into account; even if the information at the source has not changed, the traditional age metric keeps increasing linearly with time. Thus, optimizing the traditional age metric causes a portion of the resources to be wasted into some transmissions that do not contribute to the timeliness of the system. To circumvent this problem, several extended versions of the traditional age metric have been proposed \cite{yates21gossip, cho3BinaryFreshness, zhong18AoSync, maatouk20AOII, melih2020infocom, Abolhassani21version, wang19counting_process} and used in solving different problems  \cite{melih_cache_TWT, melih20LimitedCache, kaswan2021ISIT,  melih21InfectionTracking}. One such metric is the \textit{version age}, which is introduced as a measure of freshness in \cite{yates21gossip, melih2020infocom, Abolhassani21version}. 

Reference \cite{yates21gossip} uses the version age in a gossip network, where the source is updated with rate $\lambda_e$, the source updates a fully-connected network of $n$ nodes with a total update rate of $\lambda$, and each node in the network updates the remaining $n-1$ nodes with a total update rate of $\lambda$. \cite{yates21gossip} shows that the version age of an individual node in such a network scales as $O(\log n)$ with the network size $n$. Some variations of this system model have been studied in \cite{buyukates21CommunityStructure, buyukates22ClusterGossip, kaswan22slicingcoding, kaswan22jamming, kaswan22timestomp, melih2021globecom}: \cite{buyukates21CommunityStructure, buyukates22ClusterGossip} consider clustered networks with a community structure and show improvements in version age due to clustering; \cite{kaswan22slicingcoding} considers file slicing and network coding and achieves a version age of $O(1)$ for each node; \cite{kaswan22jamming, kaswan22timestomp} consider version age in the presence of adversarial attacks and investigate how adversarial actions affect the version age; and \cite{melih2021globecom} considers the binary freshness metric instead of the version age in gossiping. 

In \cite{yates21gossip}, the total gossip rate of the network is $n\lambda$. A downside of the kind of gossiping in \cite{yates21gossip} is that the nodes with staler versions also get to gossip to relatively fresher nodes, which does not actually contribute to the timeliness of the overall network. Our intuition in this paper is that, if the gossip rate of staler nodes could be assigned (shifted) to fresher nodes instead, then the timeliness of the network could be improved. The challenge is how to implement this intuition in a distributed network where there is no centralized server. 

To that end, we introduce ASUMAN, an age-aware distributed gossiping scheme. Our key idea is reminiscent of the opportunistic channel access scheme proposed in \cite{zhao5SensorNetwork, zhao5opportunistic} in a different context, different system model, with a different goal. \cite{zhao5SensorNetwork, zhao5opportunistic} consider a fading multiple access channel with distributed users. It is well-known \cite{knoppp_humblet_capacity, Tse_Hanly_98_multicaccess} that in a fading multiple access channel, in order to maximize the sum capacity, only the largest channel gain user should transmit. While the receiver may measure user channel gains and announce the largest channel gain user as a feedback in the downlink, the approach in \cite{zhao5SensorNetwork, zhao5opportunistic} is that users measure their own channel gains in the downlink, and apply an opportunistic carrier-sensing-like \cite{CSMA1975} scheme in the uplink. In \cite{zhao5SensorNetwork, zhao5opportunistic}, before starting transmissions, the users wait for a back-off time which is a decreasing function of their individual channel gains. Since the user with the largest channel gain waits the least amount of time, it starts transmitting first, all other users become aware of this, and remain silent for the duration of transmission. That is, the broadcast nature of the wireless channel is exploited as an implicit feedback mechanism for the coordination of distributed users.

We use a similar concept in the context of wireless sensor nodes with the objective of information freshness. In our setting, where each node knows its own age, we are interested in enabling the freshest node to capture the channel and update the remaining staler users. In our opportunistic gossiping scheme ASUMAN, each node waits for a back-off time proportional to its own age before starting to gossip. Since the freshest node will start gossiping first, upon hearing this, the rest of the nodes will forgo gossiping for that cycle, and will only potentially receive updates. We show that this policy achieves an age scaling of order of $O(1)$ as the number of nodes increases. For our analysis, we use the stochastic hybrid system (SHS) approach \cite{hespanha_SHS}, similar to \cite{yates21gossip}, to derive the expressions for the average steady-state age values.

\section{System Model}

\begin{figure}[t]
\centerline{\includegraphics{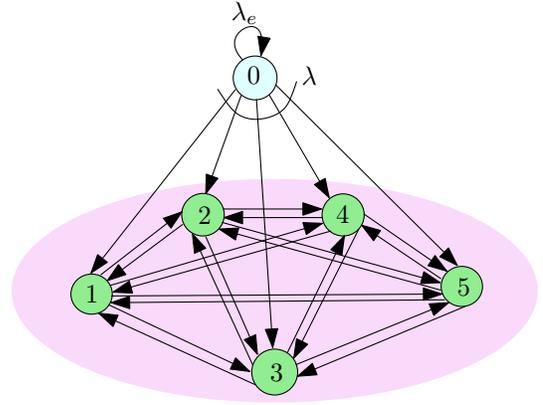}}
\caption{Source $0$ updates itself with rate $\lambda_{e}$ and sends updates to the nodes $\mathcal{N}=\{1,2,3,4,5\}$ uniformly with total rate $\lambda$, i.e., with rate $\lambda/5$ to each of the nodes. The nodes gossip with each other with total update rate $B$.}
\label{SystemModel}
\end{figure}

We consider a system with a source node, labeled as 0, and a set of nodes $\mathcal{N}=\{1,2,\ldots,n\}$; see Fig.~\ref{SystemModel}. The source node updates itself with a Poisson process with rate $\lambda_{e}$ and it sends updates to each of the nodes in the network as Poisson arrivals with rate $\frac{\lambda}{n}$. The network has a total gossiping rate $B$. The nodes gossip their knowledge about the source's information to maintain the timeliness of the overall network. We consider the version age metric for measuring this timeliness. The version age of the $i$th node, denoted as $\Delta_i(t)$, is the version of information present in the $i$th node as compared to the current version at the source. That is,
\begin{align}
    \Delta_i(t)=N_s(t)-N_i(t),    \label{Version}
\end{align}
where $N_s(t)$ is the version at the source and $N_i(t)$ is the version at the $i$th node at time $t$. We consider the age vector $\mathbf{\Delta}(t)=[\Delta_1(t), \Delta_2(t), \ldots, \Delta_n(t)]$ to denote the version of all the nodes in the network. If the source updates itself at any time, all the elements of $\mathbf{\Delta}(t)$ increase by 1. We assume that the nodes are aware of their own version age. The nodes gossip among themselves to disperse the information in the network. When node $i$ sends a gossip update to node $j$ at time $t$, node $j$ updates its information if the received information is fresher, otherwise it keeps its information as it is. Thus, the age of node $j$ is updated to $\Delta'_j(t)=\Delta_{\{i,j\}}(t)=\min \{\Delta_i(t),\Delta_j(t)\}$. 

\section{Opportunistic Gossiping via ASUMAN}

\begin{figure*}[t]
\centerline{\includegraphics[width=\textwidth]{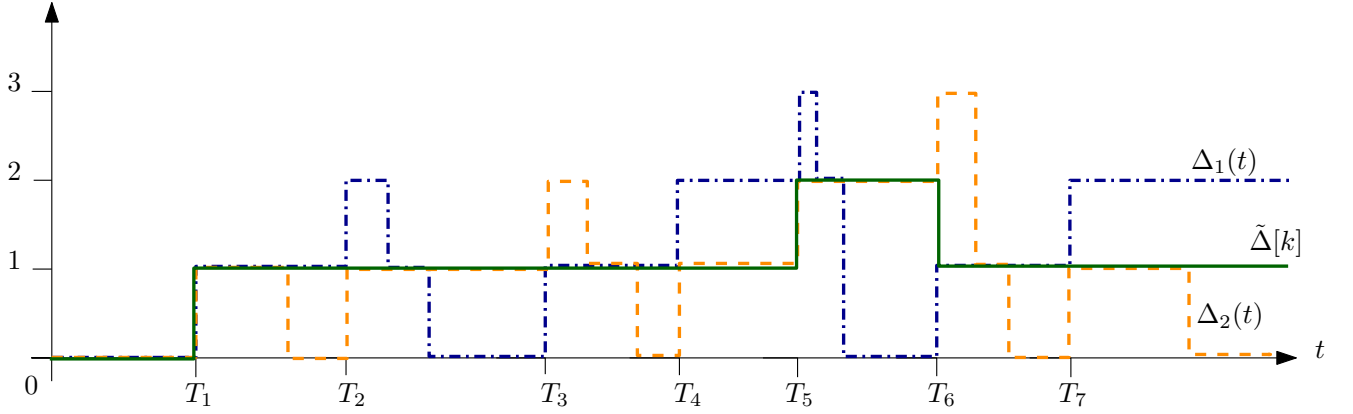}}
\caption{Source $0$ updates itself with rate $\lambda_{e}$ and updates each of the nodes $\mathcal{N}=\{1,2\}$ with rate $\lambda/2$. The version ages of the individual nodes are denoted by $\Delta_1(t)$ and $\Delta_2(t)$, respectively, and $\tilde{\Delta}[k]=\min\{\Delta_1(T_k),\Delta_2(T_k)\}$.}
\label{v_a}
\end{figure*}

In this section, we define the ASUMAN scheme and derive a theoretical upper bound for its average age of gossip. Since each node is aware of its own age, when the source updates its information, it acts as a synchronization signal for all the nodes in the network. Suppose we denote the time instances of source self updates as $T_k$, where $k$ is any positive integer. $T_0$ is defined to be $0$. The inter-arrival time $\tau_{k+1}=T_{k+1}-T_k$ are exponentially distributed with mean $\frac{1}{\lambda_{e}}$. When the source updates its information at time $T_k$, each node stops gossiping. The $i$th node waits for a time $C\Delta_i(T_k)$, where is $C$ is a small proportionality constant. After waiting this time, node $i$ broadcasts a signal to all the nodes in the network and starts gossiping. However, if a node receives a broadcast from another node before its waiting period expires, then it remains silent for the time interval $\mathcal{I}_k=[T_k,T_{k+1})$. Thus, for each time interval, only the nodes which have the lowest age at the beginning of the interval gets to gossip. At time $T_k$, we use $\mathcal{M}_k$ to denote the set of indices of the nodes with the minimum age, $\tilde{\Delta}[k]=\min_{i} \Delta_i(T_k)$; see Fig.~\ref{v_a}. 

From the broadcast signals, all the nodes in the network get to know that there are total $|\mathcal{M}_k|$ number of minimum-age nodes at $T_k$. Therefore, each of the nodes in $\mathcal{M}_k$ utilizes only $\frac{B}{|\mathcal{M}_k|}$ of total gossip rate, while all the other nodes do not use any update rate for $\mathcal{I}_k$. If $\tau_{k+1}>C\tilde{\Delta}[k]$, each node in $\mathcal{M}_k$ gossips to every other node with rate $\frac{B}{|\mathcal{M}_k|(n-1)}$ for the time interval $[T_k+C\tilde{\Delta}[k],T_{k+1})$. Otherwise, the source updates itself before the nodes get a chance to gossip opportunistically, and the next interval begins with the same scheme. In this work, we are interested in the steady-state mean of the version age of a node, which is defined as
\begin{align}
    a_i=\lim_{t\to\infty}a_i(t)=\lim_{t\to\infty}\mathbb{E}[\Delta_i(t)].\label{avg_notation}
\end{align}

To evaluate this steady-state mean age, we define the mean of $\tilde{\Delta}[k]$ as $\tilde{a}[k]=\mathbb{E}[\tilde{\Delta}[k]]$ and evaluate it in Lemma~\ref{lemma1}.

\begin{lemma}\label{lemma1}
The mean of minimum age in interval $\mathcal{I}_k$ is  
\begin{align}
\tilde{a}[k]=\sum_{\ell=0}^{k-1}\left(\frac{\lambda_{e}}{\lambda_{e}+\lambda}\right)^{\ell}, \quad k\geq 1.
\end{align}
\end{lemma}

\begin{Proof}
We use induction for the proof. Since all the ages are 0 at the beginning, $\tilde{a}[0]=0$ and $\tilde{a}[1]=1$. Assume that the statement is true for $k$. The probability that the source does not update any node in $\mathcal{I}_k$ is $e^{-\lambda\tau_{k+1}}$. If any node in the network is updated in $\mathcal{I}_k$, $\tilde{\Delta}[k+1]$ becomes 1; otherwise it is $\tilde{\Delta}[k]+1$. Thus, we have
\begin{align}
    \tilde{a}[k+1]=\mathbb{E}[(1-e^{-\lambda \tau_{k+1}})+(\tilde{\Delta}[k]+1)e^{-\lambda \tau_{k+1}}].
\end{align}
Since $\tau_{k+1}$ is exponentially distributed with parameter $\lambda_{e}$,
\begin{align}
    \mathbb{E}\left[e^{-\lambda \tau_{k+1}}\right]&=\int_0^{\infty}e^{-\lambda \tau_{k+1}}\lambda_{e}e^{-\lambda_{e} \tau_{k+1}}d\tau_{k+1}\\
    &=\frac{\lambda_{e}}{\lambda_{e}+\lambda}.
\end{align}
Thus, we obtain
\begin{align}
    \tilde{a}[k+1]=1+\tilde{a}[k]\frac{\lambda_{e}}{\lambda_{e}+\lambda}.\label{recurrence}
\end{align}
Now, using the induction hypothesis, we can rewrite \eqref{recurrence} as
\begin{align}
    \tilde{a}[k+1]&=1+\sum_{\ell=0}^{k-1}\left(\frac{\lambda_{e}}{\lambda_{e}+\lambda}\right)^{\ell}\cdot\frac{\lambda_{e}}{\lambda_{e}+\lambda}\\
    &=\sum_{\ell=0}^{k}\left(\frac{\lambda_{e}}{\lambda_{e}+\lambda}\right)^{\ell},
\end{align}
completing the proof.
\end{Proof}

Next, in Lemma~\ref{lemma2}, we consider an unrealistic case of $C=0$, i.e., all the nodes instantaneously know about the minimum age nodes in the beginning of the interval $\mathcal{I}_k$. Although this is not a feasible model, the result of this lemma will be used for calculations in the case of $C>0$. In addition, in Lemma~\ref{lemma2}, we consider the case where the total gossip  rate of the network is $B=n \lambda$, which is the same as the total gossip rate in \cite{yates21gossip}.  

\begin{lemma}\label{lemma2}
For $C = 0$, if the total gossip rate is $B=n\lambda$, then the steady-state mean version age of a node scales as $O(1)$.
\end{lemma}

\begin{Proof}
Since the system is symmetric with respect to any node in the network, proving the result only for any fixed $i$th node will suffice. To analyze the system, we follow the SHS formulation in \cite{hespanha_SHS}. Since $C=0$, only one type of state transition is involved. Thus, $\mathcal{Q}=\{0\}$ and for the node $i\in\mathcal{N}$, we choose a function $\psi_i:\mathbbm{R}^n\times[0,\infty)\to\mathbbm{R}$, such that
\begin{align}
    \psi_i(\mathbf{\Delta}(t),t)=\Delta_i(t).   
\end{align}
Following \cite[Thm.~1]{hespanha_SHS}, we write the expected value of the extended generator function as
\begin{align}
    \mathbb{E}[(L\psi_i)(\mathbf{\Delta}(t),t)]=&\sum_{(j,\ell)\in \mathcal{L}}\lambda_{j,\ell}(\mathbf{\Delta}(t),t)\mathbb{E}[\psi_i(\phi_{j,\ell}(\mathbf{\Delta}(t),t))\notag\\
    &\qquad-\psi_i(\mathbf{\Delta}(t),t)],
\end{align}
where $\mathcal{L}$ is the set of all possible state transitions. Define reset maps $\phi_{j,\ell}(\mathbf{\Delta}(t),t)=\hat{\mathbf{\Delta}}(t)=[\hat{\Delta}_1(t),\hat{\Delta}_n(t),\ldots,\hat{\Delta}_n(t)]$ as
\begin{align}
    \hat{\Delta}_i(t)=
    \left\{
	\begin{array}{ll}
		\Delta_i(t)+1,  &\mbox{if } j = 0, \ell=0 \\
		0, &\mbox{if } j=0, \ell=i\\
		\min(\Delta_j(t),\Delta_\ell(t)),  & \mbox{if } j\in\mathcal{N},\ell=i \\
		\Delta_i(t), &\mbox{otherwise}.
	\end{array}
\right.
\end{align}
The update rates $\lambda_{j,\ell}$ are given as
\begin{align}
    \lambda_{j,\ell}(\mathbf{\Delta}(t),t)=
    \left\{
	\begin{array}{ll}
		\lambda_{e}, &\mbox{if } j = 0,\ell=0 \\
		\frac{\lambda}{n}, &\mbox{if } j=0,\ell=i\\
		\lambda_{j,\ell}^{(k)}(t), &\mbox{otherwise},
	\end{array}
\right.
\end{align}
where $\lambda_{j,\ell}^{(k)}(t)$ is the gossip rate of node $j$ to node $l$ in the time interval $\mathcal{I}_k$. Since $C=0$,
\begin{align}
    \lambda_{j,\ell}^{(k)}(t)=
    \left\{
	\begin{array}{ll}
		\frac{B}{|\mathcal{M}_k|(n-1)}, &\mbox{if } j\in\mathcal{M}_k, \ell\in\mathcal{N}, t\in\mathcal{I}_k \\
		0, &\mbox{otherwise.}
	\end{array}
\right.\label{GossipRate1}
\end{align}
Using the notations introduced in \eqref{avg_notation}, we rewrite the expected value of the extended generator function as
\begin{align}
    \mathbb{E}&[(L\psi_{i})(\mathbf{\Delta}(t),t)]\notag\\
    &=\mathbb{E}\bigg[\lambda_{e}(\Delta_i(t)+1-\Delta_i(t))+\frac{\lambda}{n}(0-\Delta_i(t))\notag\\
    &\ \ \ +\sum_{j\in \mathcal{N}}\lambda_{j,i}(\mathbf{\Delta}(t),t)\left(\Delta_{\{j,i\}}^{(k)}(t)-\Delta_i(t)\right)\bigg].
\end{align}
Now, for $t\in\mathcal{I}_k$, we write the expectation as 
\begin{align}
   &\!\!\!\!\mathbb{E}[(L\psi_{i})(\mathbf{\Delta}(t),t)]\notag\\
    &\!\!\!\!=\lambda_{e}-\frac{\lambda}{n}a_i(t)+\mathbb{E}\bigg[\sum_{j\in \mathcal{M}_k}\lambda_{j,i}^{(k)}(t)(\Delta_{\{j,i\}}^{(k)}(t)-\Delta_i(t))\bigg].\label{GossipEquation}
\end{align}

\begin{figure*}[t]
\centerline{\includegraphics[width=0.94\textwidth]{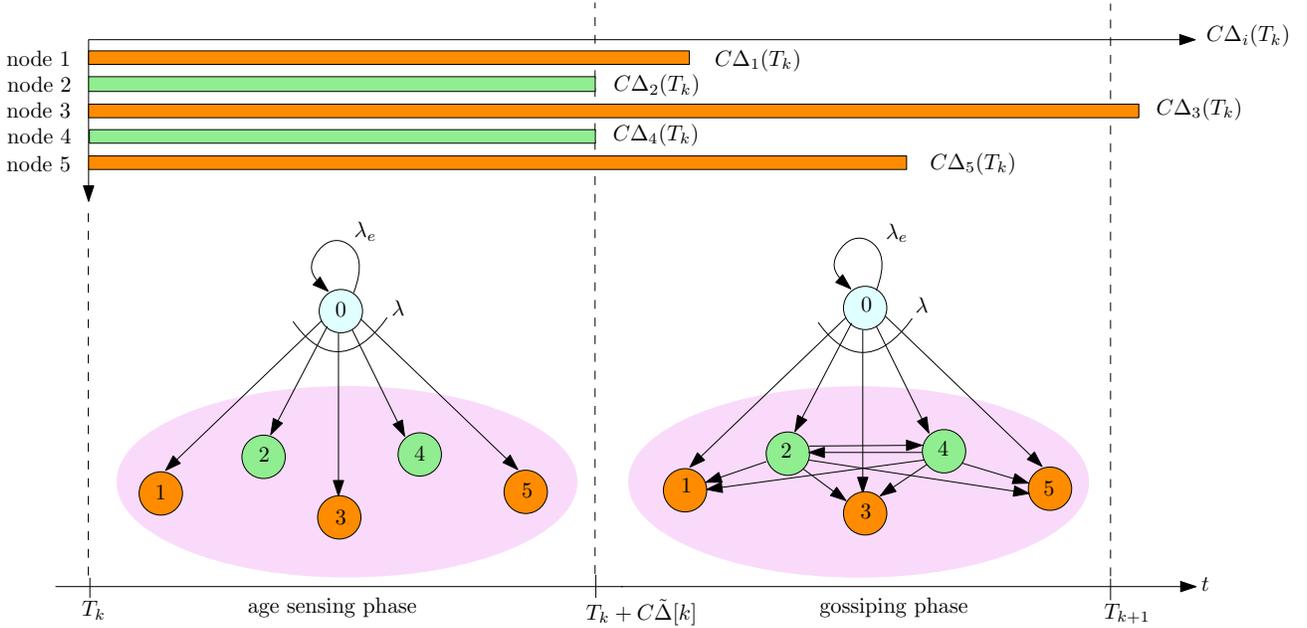}}
\caption{An example of a typical opportunistic gossiping in a 5 node network. At $T_k$, the minimum age nodes are $\mathcal{M}_k=\{2,4\}$. Thus, they wait for a time of $\tilde{\Delta}[k]$ in the age sensing phase $[T_k, T_k+C\tilde{\Delta}[k])$ and start transmitting with total update rate $B$ in the gossiping phase $[T_k+C\tilde{\Delta}[k],T_{k+1})$.}
\label{illustration}
\end{figure*}

Note that, \eqref{GossipRate1} is true, if the $i$th node is not in $\mathcal{M}_k$ by the formulation of our proposed gossiping scheme. However, even if node $i$ is in $\mathcal{M}_k$ we can still assume that it is gossiping to itself with rate $\frac{B}{|\mathcal{M}_k|(n-1)}$, since the corresponding product term $(\Delta_{\{i,i\}}^{(k)}(t)-\Delta_i(t))=0$. Now, since the version age is a piece-wise constant function of time, we obtain
\begin{align}
    \frac{d\mathbb{E}[\psi_i(\mathbf{\Delta}(t),t)]}{dt}=\frac{d\mathbb{E}[\Delta_i(t)]}{dt}=0
\end{align}
for all the continuity points. Hence, the expected value in \eqref{GossipEquation} is $0$, by Dynkin’s formula, as given in \cite{hespanha_SHS}. Thus, \eqref{GossipEquation} becomes
\begin{align}
    0=&\:\lambda_{e}-\frac{\lambda}{n}a_i(t)\notag\\
    &+\mathbb{E}\bigg[\sum_{j\in \mathcal{M}_k}\frac{B}{|\mathcal{M}_k|(n-1)}\left(\Delta_{\{j,i\}}^{(k)}(t)-\Delta_i(t)\right)\bigg].
\end{align}
Hence, the mean age of an individual node is expressed as
\begin{align}
    &\left(\frac{\lambda}{n}+\frac{B}{n-1}\right)a_i(t)\notag\\
    &\qquad =\lambda_e+\mathbb{E}\bigg[\sum_{j\in \mathcal{M}_k}\frac{B}{|\mathcal{M}_k|(n-1)}\Delta_{\{j,i\}}^{(k)}(t)\bigg].\label{j_dep}
\end{align}
In \eqref{j_dep}, $\mathcal{M}_k$ is a function of $\mathbf{\Delta}(t)$. Instead of deriving the distribution of $\mathbf{\Delta}(t)$, we use the inequality $\Delta_{\{j,i\}}^{(k)}(t)\leq \tilde{\Delta}[k]$ for $t\in\mathcal{I}_k$, and rewrite \eqref{j_dep} as the following upper bound 
\begin{align}
    a_i(t)&\leq\frac{\lambda_{e}+\frac{B}{n-1}\mathbb{E}\left[\sum_{j\in \mathcal{M}_k}\frac{\tilde{\Delta}[k]}{|\mathcal{M}_k|}\right]}{\frac{\lambda}{n}+\frac{B}{n-1}},\quad\forall t \in \mathcal{I}_k\\
    &=\frac{\lambda_{e}+\frac{B}{n-1}\tilde{a}[k]}{\frac{\lambda}{n}+\frac{B}{n-1}},\quad\forall t \in \mathcal{I}_k.\label{UB1}
\end{align}
We are interested in the steady-state average age, i.e., average age at $t\to\infty$. We evaluate the asymptote of the upper bound in \eqref{UB1} as $k\to\infty$. From Lemma~\ref{lemma1}, we have 
\begin{align}
    \lim_{k\to \infty}\tilde{a}[k]=\frac{\lambda_{e}+\lambda}{\lambda}.\label{steady_state}
\end{align}

Using this result in \eqref{UB1}, we obtain
\begin{align}
    a_i=\lim_{t\to\infty}a_i(t)&\leq\lim_{k\to\infty}\frac{\lambda_{e}+\frac{B}{n-1}\tilde{a}[k]}{\frac{\lambda}{n}+\frac{B}{n-1}}\\
    &=\frac{\lambda_{e}}{\lambda}\frac{\left(1+\frac{B}{n-1}\left(\frac{1}{\lambda}+\frac{1}{\lambda_{e}}\right)\right)}{\left(\frac{1}{n}+\frac{n}{n-1}\right)}.\label{UB2}
\end{align}
Now, to calculate the scaling of the average age, we use the relation that $B=n\lambda$, which yields
\begin{align}
    \lim_{n\to\infty}a_i&\leq\lim_{n\to\infty}\frac{\lambda_{e}}{\lambda}\frac{(1+\frac{n\lambda}{n-1}(\frac{1}{\lambda}+\frac{1}{\lambda_{e}}))}{\left(\frac{1}{n}+\frac{n}{n-1}\right)}\\
    &=\frac{2\lambda_{e}}{\lambda}+1,
\end{align}
concluding the proof.
\end{Proof} 

Now, we use the results of Lemmas~\ref{lemma1} and \ref{lemma2}, to formulate the average version age in Theorem~\ref{thm1}.

\begin{theorem}\label{thm1}
For $C>0$, if $C$ is chosen such that it is bounded for all $n$ and $C\to 0$ as $n\to\infty$, keeping the total gossip rate the same as before, i.e., $B=n\lambda$, then the average version age of a node scales as $O(1)$, and the asymptotic upper bound is the same as that for the case of $C=0$.   
\end{theorem}

\begin{Proof}
There is no change in the function $\psi_i$ or in the reset maps $\phi_{j,l}$. The only change is in the update frequencies. For any choice of $C>0$, we divide the time interval $\mathcal{I}_k$ into two phases, an age sensing phase $\mathcal{I}^{(s)}_k=[T_k,\min(T_k+C\tilde{\Delta}[k],T_{k+1}))$ and a gossiping phase $\mathcal{I}^{(g)}_k=[\min(T_k+C\tilde{\Delta}[k],T_{k+1}),T_{k+1})$; see Fig.~\ref{illustration}. 

We already have an upper bound for the steady-state average age expression for $\mathcal{I}^{(g)}_k$ from Lemma~\ref{lemma2}. Let us denote the right hand side in \eqref{UB1} as $a^{(g)}[k]$, i.e., 
\begin{align}
a^{(g)}[k]=\frac{\lambda_{e}+\frac{B}{n-1}\tilde{a}[k]}{\frac{\lambda}{n}+\frac{B}{n-1}}.
\end{align}

For $\mathcal{I}^{(s)}_k$, we evaluate an upper bound by ignoring the source to $i$th node updates and only considering the opportunistic gossiping. We define the process $\Delta^{(s)}[k]$, such that $\Delta^{(s)}[1]=1$. If the $(k-1)$th interval does not have a gossiping phase, i.e., $\tau_k\leq C\tilde{\Delta}[k-1]$, or if none of the active nodes in $\mathcal{M}_{k-1}$ gossip to node $i$ in $\mathcal{I}^{(g)}_{k-1}$, then $\Delta^{(s)}[k]=\Delta^{(s)}[k-1]+1$. Otherwise, if any node in $\mathcal{M}_{k-1}$ gossips to node $i$ in the interval $\mathcal{I}^{(g)}[k]$, then $\Delta^{(s)}[k]=\tilde{\Delta}[k-1]+1$. We express the probabilistic recurrence relations for $k>1$ as
\begin{align}
    \Delta^{(s)}[k]=
\left\{
	\begin{array}{ll}
		\Delta^{(s)}[k-1]+1, &\mathbb{P}(\tau_{k}\leq C\tilde{\Delta}[k-1])\\
		\Delta^{(s)}[k-1]+1,  &\mathbb{P}(\tau_k > C\tilde{\Delta}[k-1])\\
		&\times e^{-\frac{B}{n-1}(\tau_k-C\tilde{\Delta}[k-1])}\\
		\tilde{\Delta}[k-1]+1, &\mathbb{P}(\tau_k > C\tilde{\Delta}[k-1])\\
		&\times \left(1-e^{-\frac{B}{n-1}(\tau_k-C\tilde{\Delta}[k-1])}\right)
	\end{array}
\right.\label{delta_s}
\end{align}
Clearly, $\Delta_i(t)\leq\Delta^{(s)}[k]$ for $t\in\mathcal{I}^{(s)}_k$. We write the mean of this upper bound as
\begin{align}
    a^{(s)}[k]=&\: \mathbb{E}[\Delta^{(s)}[k]]\notag\\
     = &\:\mathbb{E}\bigg[(\Delta^{(s)}[k-1]+1)\bigg(\mathbb{P}(\tau_k\leq C\tilde{\Delta}[k-1])\notag\\
    & \ \ \ +\mathbb{P}(\tau_k > C\tilde{\Delta}[k-1])e^{-\frac{B}{n-1}(\tau_k-C\tilde{\Delta}[k-1])}\bigg)\notag\\
    & +(\tilde{\Delta}[k-1]+1)\mathbb{P}(\tau_k > C\tilde{\Delta}[k-1])\notag\\
    & \ \ \ \times\left(1-e^{-\frac{B}{n-1}(\tau_k-C\tilde{\Delta}[k-1])}\right)\bigg].\label{SensinUB0}
\end{align}
Since $\tau_k$ is exponentially distributed, we rewrite \eqref{SensinUB0} as
\begin{align}
    a^{(s)}[k]=&\: (a^{(s)}[k-1]+1)\mathbb{E}\big[1-e^{-\lambda_e C\tilde{\Delta}[k-1]}\notag\\
    & \ \ +e^{-\lambda_e C\tilde{\Delta}[k-1]}\cdot e^{-\frac{B}{n-1}(\tau_k-C\tilde{\Delta}[k-1])}\big]\notag\\
    & +(\tilde{a}[k-1]+1)\mathbb{E}\bigg[e^{-\lambda_e C\tilde{\Delta}[k-1]}\notag\\
    & \ \ \times\left(1-e^{-\frac{B}{n-1}(\tau_k-C\tilde{\Delta}[k-1])}\right)\bigg].\label{SensingUB}
\end{align}
Now, for $B=n\lambda$, as $n\to \infty$, $C\to 0$, $\frac{B}{n-1}\to \lambda$, $e^{-\lambda_e C\tilde{\Delta}[k-1]}\to 1$, and $e^{\frac{B}{n-1} C\tilde{\Delta}[k-1]}\to 1$. Thus, \eqref{SensingUB} becomes 
\begin{align}
    \lim_{n\to\infty}a^{(s)}[k]=&\left(\lim_{n\to\infty}a^{(s)}[k-1]+1\right)\mathbb{E}\left[e^{-\lambda\tau_k}\right]\notag\\
    &+\left(\tilde{a}[k-1]+1\right)\mathbb{E}\left[1-e^{-\lambda\tau_k}\right]. \label{recursive_eqn}
\end{align}
From Lemma~\ref{lemma1}, we know that $\mathbb{E}\left[e^{-\lambda\tau_k}\right]=\frac{\lambda_e}{\lambda_e+\lambda}$. Lemma~\ref{lemma1} also says
\begin{align}
    \tilde{a}[k-1]=\sum_{\ell=0}^{k-2}\left(\frac{\lambda_{e}}{\lambda_{e}+\lambda}\right)^{\ell}&\leq \lim_{k\to\infty}\sum_{\ell=0}^{k-1}\left(\frac{\lambda_{e}}{\lambda_{e}+\lambda}\right)^{\ell}\\
    &=\frac{\lambda_e+\lambda}{\lambda}.\label{a_tilde_UB}
\end{align}
Using \eqref{a_tilde_UB} in \eqref{recursive_eqn} gives
\begin{align}
    \lim_{n\to\infty}a^{(s)}[k]\leq 2+\frac{\lambda_e}{\lambda_e+\lambda}\lim_{n\to\infty}a^{(s)}[k-1]. \label{recursive_eqn2}
\end{align}

Here, we define a new sequence $b[k]$, such that $b[1]=\lim_{n\to\infty}a^{(s)}[1]=1$, and evolves according to
\begin{align}
    b[k]=2+\frac{\lambda_e}{\lambda_e+\lambda}b[k-1]. \label{recursive_eqn3}
\end{align}
Therefore, $b[k]\geq \lim_{n\to\infty}a^{(s)}[k]$ for all $k$. Now, using similar logic as in Lemma~\ref{lemma1} here, we obtain an expression for $b[k]$ as 
\begin{align}
    b[k]=2\sum_{\ell=0}^{k-2}\left(\frac{\lambda_{e}}{\lambda_{e}+\lambda}\right)^{\ell}+\left(\frac{\lambda_{e}}{\lambda_{e}+\lambda}\right)^{k-1}.\label{b_k}
\end{align}
Hence, we obtain the relation 
\begin{align}
    \lim_{n\to\infty}a^{(s)}[k]\leq 2\sum_{\ell=0}^{k-2}\left(\frac{\lambda_{e}}{\lambda_{e}+\lambda}\right)^{\ell}+\left(\frac{\lambda_{e}}{\lambda_{e}+\lambda}\right)^{k-1}.\label{a_k}
\end{align}

From \eqref{a_k}, we conclude that $a^{(s)}[k]\sim O(1)$. Note that, since $\Delta_i(t)\leq\Delta^{(s)}[k]$ for $t\in\mathcal{I}_k^{(s)}$, the gossiping process in $\mathcal{I}_k^{(g)}$ cannot increase the age. Thus, $\Delta_i(t)\leq\Delta^{(s)}[k]$ for all $t\in\mathcal{I}_k$. Therefore, $a_{i}(t)\sim O(1)$. This finishes the first part of the statement of Theorem~\ref{thm1}, which is that the asymptotic upper bound for the average age is $O(1)$. 

To prove the next part of the theorem, i.e., that under the conditions given in the statement of the theorem, the upper for $C>0$ is the same as the upper bound for $C=0$, first we note that \eqref{a_k} yields the following steady-state upper bound
\begin{align}
    \!\!\!\lim_{k\to\infty}2\sum_{\ell=0}^{k-2}\left(\frac{\lambda_{e}}{\lambda_{e}+\lambda}\right)^{\ell}+\left(\frac{\lambda_{e}}{\lambda_{e}+\lambda}\right)^{k-1}\!\!=2\left(\frac{\lambda_e}{\lambda}+1\right). \!\!
\end{align}
Comparing to the case of $C=0$, we see that this value is $\lim_{k\to\infty}a^{(g)}[k]+1$. To get a tighter upper bound, we take the age reduction in the gossiping phase into consideration. So far, we have
\begin{align}
    a_i(t)\leq
\left\{
	\begin{array}{ll}
		a^{(s)}[k],  &\text{if } \tau_{k+1}\leq C\tilde{\Delta}[k] \ \forall t \in \mathcal{I}_k\\
		a^{(s)}[k],  &\text{if } \tau_{k+1}> C\tilde{\Delta}[k] \ \forall t \in \mathcal{I}^{(s)}_k\\
		a^{(g)}[k], &\text{if } \tau_{k+1}> C\tilde{\Delta}[k] \ \forall t \in \mathcal{I}^{(g)}_k
	\end{array}
\right.\label{ineq}
\end{align}
We calculate the average age as
\begin{align}
    a_i=\lim_{T\to\infty}\frac{1}{T}\int_{0}^{T}\Delta_i(t)dt=\lim_{T\to\infty}\frac{1}{T}\sum_{k=1}^{N(T)}\beta_i[k], \label{avg_age}
\end{align}
where we denote the number of source self updates as $N(T)=\max\{j:T_j\leq T\}$ and $\beta_i[k]=\int_{\mathcal{I}_k}\Delta_i(t)dt$. Assuming ergodicity of the process, we rewrite \eqref{avg_age} as
\begin{align}
    a_i=\lim_{T\to\infty}\frac{\frac{1}{N(T)}\sum_{k=1}^{N(T)}\beta_i[k]}{T/N(T)}=\frac{\lim_{k\to\infty}\mathbb{E}\left[\beta_i[k]\right]}{\lim_{T\to\infty}T/N(T)}, \label{avg_age2}
\end{align}
if $\lim_{k\to\infty}\mathbb{E}\left[\beta_i[k]\right]$ converges. Since the source self update is a Poisson process with rate $\lambda_e$, $\lim_{T\to\infty}T/N(T)=\frac{1}{\lambda_e}$. We write the numerator of \eqref{avg_age2} as
\begin{align}
    \mathbb{E}\left[\beta_i[k]\right]=&\:\mathbb{E}\left[\int_{\mathcal{I}_k}\Delta_i(t)\mathbbm{1}\{\tau_{k+1}\leq C\tilde{\Delta}[k]\}dt\right]\notag\\
    &+\mathbb{E}\left[\int_{\mathcal{I}_k}\Delta_i(t)\mathbbm{1}\{\tau_{k+1}> C\tilde{\Delta}[k]\}dt\right],\label{avg_age3}
\end{align}
where $\mathbbm{1}\{\cdot\}$ is the indicator function. The first term in \eqref{avg_age3} constitutes the event when the source updates too quickly for the network to get into the gossiping phase. It is bounded as
\begin{align}
    &\mathbb{E}\left[\int_{\mathcal{I}_k}\Delta_i(t)\mathbbm{1}\{\tau_{k+1}\leq C\tilde{\Delta}[k]\}dt\right]\notag\\
    &=\mathbb{E}_{\tilde{\Delta}[k]}\left[\mathbb{E}\left[\int_{\mathcal{I}_k}\Delta_i(t)\mathbbm{1}\{\tau_{k+1}\leq C\tilde{\Delta}[k]\}dt\bigg|\tilde{\Delta}[k]\right]\right]\\
    &\leq\mathbb{E}_{\tilde{\Delta}[k]}\left[a^{(s)}[k]\mathbb{E}_{\tau_{k+1}}\left[\tau_{k+1}\mathbbm{1}\{\tau_{k+1}\leq C\tilde{\Delta}[k]\}\bigg|\tilde{\Delta}[k]\right]\right].\label{term1}
\end{align}
We obtain the inner expectation in \eqref{term1} as 
\begin{align}
    &\mathbb{E}_{\tau_{k+1}}\left[\tau_{k+1}\mathbbm{1}\{\tau_{k+1}\leq C\tilde{\Delta}[k]\}\bigg|\tilde{\Delta}[k]\right]\notag\\
    &\quad=\int_{0}^{C\tilde{\Delta}[k]}\tau_{k+1}\lambda_e e^{-\lambda_e\tau_{k+1}}d\tau_{k+1}\\
    &\quad=\frac{1}{\lambda_e}\left(1-e^{-\lambda_e C\tilde{\Delta}[k]}(\lambda_e C\tilde{\Delta}[k] + 1)\right).
\end{align}
For the second term in \eqref{avg_age3}, we break the integral into age sensing and gossiping phases as follows
\begin{align}
    &\mathbb{E}\left[\int_{\mathcal{I}_k}\Delta_i(t)\mathbbm{1}\{\tau_{k+1}> C\tilde{\Delta}[k]\}dt\right]\notag\\
    &=\mathbb{E}_{\tilde{\Delta}[k]}\bigg[\mathbb{E}\bigg[\int_{\mathcal{I}^{(s)}_k}\Delta_i(t)\mathbbm{1}\{\tau_{k+1}> C\tilde{\Delta}[k]\}dt \notag\\
    &\ \ \ +\int_{\mathcal{I}^{(g)}_k}\Delta_i(t)\mathbbm{1}\{\tau_{k+1}> C\tilde{\Delta}[k]\}dt\bigg|\tilde{\Delta}[k]\bigg]\bigg]\\
    &\leq \mathbb{E}_{\tilde{\Delta}[k]}\bigg[a^{(s)}[k]\mathbb{E}_{\tau_{k+1}}\bigg[C\tilde{\Delta}[k]\mathbbm{1}\{\tau_{k+1}> C\tilde{\Delta}[k]\}\bigg|\tilde{\Delta}[k]\bigg]\bigg] \notag\\
    &\ \ \ +\mathbb{E}_{\tilde{\Delta}[k]}\bigg[a^{(g)}[k]\mathbb{E}_{\tau_{k+1}}\bigg[(\tau_{k+1}-C\tilde{\Delta}[k])\notag\\
    &\ \ \ \times\mathbbm{1}\{\tau_{k+1}> C\tilde{\Delta}[k]\}\bigg|\tilde{\Delta}[k]\bigg]\bigg]\\
    &=\mathbb{E}_{\tilde{\Delta}[k]}\bigg[(a^{(s)}[k]-a^{(g)}[k])\notag\\
    &\ \ \ \times\mathbb{E}_{\tau_{k+1}}\bigg[C\tilde{\Delta}[k]\mathbbm{1}\{\tau_{k+1}> C\tilde{\Delta}[k]\}\bigg|\tilde{\Delta}[k]\bigg]\bigg]\notag\\
    &\ \ \ +\mathbb{E}_{\tilde{\Delta}[k]}\bigg[a^{(g)}[k]\mathbb{E}_{\tau_{k+1}}\bigg[\tau_{k+1}\mathbbm{1}\{\tau_{k+1}> C\tilde{\Delta}[k]\}\bigg|\tilde{\Delta}[k]\bigg]\bigg].
\end{align}
We evaluate the inner expectations as 
\begin{align}
    &\mathbb{E}_{\tau_{k+1}}\bigg[C\tilde{\Delta}[k]\mathbbm{1}\{\tau_{k+1}> C\tilde{\Delta}[k]\}\bigg|\tilde{\Delta}[k]\bigg]\notag\\
    &\quad=C\tilde{\Delta}[k]\mathbb{P}(\tau_{k+1}> C\tilde{\Delta}[k])=C\tilde{\Delta}[k]e^{-\lambda_eC\tilde{\Delta}[k]},
\end{align}
and
\begin{align}
    &\mathbb{E}_{\tau_{k+1}}\left[\tau_{k+1}\mathbbm{1}\{\tau_{k+1}> C\tilde{\Delta}[k]\}\bigg|\tilde{\Delta}[k]\right]\notag\\
    &\quad=\int_{C\tilde{\Delta}[k]}^{\infty}\tau_{k+1}\lambda_e e^{-\lambda_e\tau_{k+1}}d\tau_{k+1}\\
    &\quad=\frac{1}{\lambda_e}e^{-\lambda_e C\tilde{\Delta}[k]}(\lambda_e C\tilde{\Delta}[k] + 1).
\end{align}
Therefore, we rewrite \eqref{avg_age3} as
\begin{align}
    &\mathbb{E}\left[\beta_i[k]\right]\notag\\
    &\quad\leq\mathbb{E}_{\tilde{\Delta}[k]}\bigg[\underbrace{\frac{a^{(s)}[k]}{\lambda_e}\left(1-e^{-\lambda_e C\tilde{\Delta}[k]}(\lambda_e C\tilde{\Delta}[k] + 1)\right)}_{\text{bounded quantity}}\notag\\
    &\quad \quad+\underbrace{(a^{(s)}[k]-a^{(g)}[k])C\tilde{\Delta}[k]e^{-\lambda_eC\tilde{\Delta}[k]}}_{\text{bounded quantity}}\notag\\
    &\quad \quad +\underbrace{\frac{a^{(g)}[k]}{\lambda_e}e^{-\lambda_e C\tilde{\Delta}[k]}(\lambda_e C\tilde{\Delta}[k] + 1)}_{\text{bounded quantity}}\bigg].\label{bounded}
\end{align}

Now, we evaluate the asymptotic scaling $\lim_{n\to\infty}\mathbb{E}\left[\beta_i[k]\right]$. Since all the age metrics $a^{(s)}[k]$, $a^{(g)}[k]$ and $\tilde{\Delta}[k]$ on the right hand side of \eqref{bounded} are upper bounded by $k$ for all values of $n$, and $C$ is bounded, we use the bounded convergence theorem to exchange the limit and expectation to calculate its scaling as $n$ becomes large. From \eqref{a_k}, it is evident that $a^{(s)}[k]\sim O(1)$. From Lemma~\ref{lemma1}, we have that $a^{(g)}[k]\sim O(1)$. As $n\to\infty,$ the quantity $\lambda_{e}C\tilde{\Delta}[k]\to 0$. Thus, we have
\begin{align}
    \lim_{n\to\infty}\mathbb{E}\left[\beta_i[k]\right]&\leq\mathbb{E}_{\tilde{\Delta}[k]}\left[\lim_{n\to\infty}\frac{a^{(g)}[k]}{\lambda_e}\right]\\
    &=\mathbb{E}_{\tilde{\Delta}[k]}\left[\lim_{n\to\infty}\frac{\lambda_{e}+\frac{B}{n-1}\tilde{a}[k]}{\lambda_e\left(\frac{\lambda}{n}+\frac{B}{n-1}\right)}\right]\\
    &=\mathbb{E}_{\tilde{\Delta}[k]}\left[\frac{\lambda_{e}+\lambda \tilde{a}[k]}{\lambda_e\lambda}\right]=\frac{\lambda_{e}+\lambda \tilde{a}[k]}{\lambda_e\lambda}.\label{scaling}
\end{align}
Therefore, using \eqref{scaling} in \eqref{avg_age2}, we obtain the asymptotic upper-bound for the average version age as
\begin{align}
    \lim_{n\to\infty}a_i\leq\lim_{k\to\infty}\lambda_e\times\frac{\lambda_{e}+\lambda \tilde{a}[k]}{\lambda_e\lambda}=\frac{2\lambda_{e}}{\lambda}+1,\label{final_UB}
\end{align}
concluding the proof.
\end{Proof}

\section{Numerical Results}

In this section, we compare our analytically derived results to numerical simulations. We choose $C=\frac{1}{n}$ for the simulations and calculate the average version age of a single node for up to $n=600$ nodes. We use $\frac{\lambda_{e}}{\lambda}=1$ and $\frac{\lambda_{e}}{\lambda}=2$ for the calculations. We also simulate the average version age from \cite{yates21gossip} gossiping policy as a comparison. 

\begin{figure}[t]
\centerline{\includegraphics[scale=0.65]{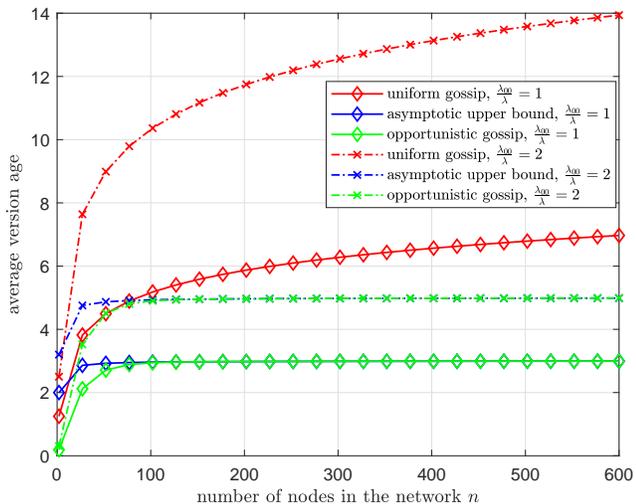}}
\caption{Average version age of a single node versus the total number of nodes in the network $n$.}
\label{plot}
\end{figure}

The results of the simulations are shown in Fig.~\ref{plot}. From Fig.~\ref{plot}, it is evident that the opportunistic gossiping of ASUMAN performs better than uniform gossiping. The uniform gossip average age scales as $O(\log n)$, whereas the asymptotic upper bound for the average age in opportunistic gossiping scales as $O(1)$ as proven in Theorem~\ref{thm1}. As calculated from \eqref{final_UB}, the upper bound is 3 and 5, for $\frac{\lambda_{e}}{\lambda}=1$ and $\frac{\lambda_{e}}{\lambda}=2$, respectively. The simulations show that the upper bound is loose when $n$ is small, and it gets tighter as $n$ becomes large. This is expected because the overall network is being updated from the source with rate $\lambda$. Therefore, with large $n$, the update rate of each individual node $\frac{\lambda}{n}$ gets smaller. Hence, in the interval $\mathcal{I}_k$, only a few nodes get updated directly from the source. However, for small $n$, the number of such nodes will be higher. This results in the average node age to be lower than the upper bound in \eqref{final_UB}. Also, we notice that the asymptotic upper bound is an increasing function of $\frac{\lambda_{e}}{\lambda}$. This result matches intuition. If $\frac{\lambda_{e}}{\lambda}$ increases, that means that the source is updating itself more frequently as compared to updating the network. This would result in higher average age. The opposite effect happens when $\lambda$ increases instead of $\lambda_{e}$, thus, resulting in lower average age.

\section{Conclusion}
We proposed ASUMAN, a gossiping policy for a network of nodes, where the nodes gossip opportunistically instead of uniformly. The network gets synchronized when the source updates itself, and the fresher nodes of the network enter into gossiping phase, following an age sensing phase. This policy allows nodes with relatively fresher versions to gossip with higher rates and nodes with staler versions to remain silent. We showed that in dense networks, the average age of a node for such a system scales as $O(1)$, which is an improvement compared to gossiping uniformly, where the average version age of a node scales as $O(\log n)$.

\bibliographystyle{unsrt}
\bibliography{reference}

\end{document}